\documentclass[prl,reprint,superscriptaddress,longbibliography]{revtex4-2}
\usepackage{braket}
\usepackage{amsmath}
\usepackage{graphicx,overpic}
\usepackage{natbib}
\usepackage[section]{placeins}
\usepackage[breaklinks]{hyperref}
\usepackage{amssymb}
\usepackage{amsmath,amsthm,bm}
\usepackage{amsfonts}%,dsfont}
\usepackage{cleveref}
\usepackage{color,soul} %just for temporary highlighting remove before submission
\hypersetup{citecolor=red,colorlinks=true,urlcolor=blue}

\usepackage[cal=stixtwofancy]{mathalpha}
\usepackage{pgfplots,amsfonts,mathtools}
\usepackage{amsfonts,amsmath,amssymb,bm}
\usepackage{graphicx,color,url,orcidlink}
\usepackage[english]{babel}

\begin{document} 

\title{Neural Quantum Propagators for Driven-Dissipative Quantum Dynamics}

\author{Jiaji Zhang\,\orcidlink{0000-0003-2978-274X}}
\affiliation{Zhejiang Laboratory, Hangzhou 311100, China}

\author{Carlos L. Benavides-Riveros\,\orcidlink{0000-0001-6924-727X}}
\email{cl.benavidesriveros@unitn.it}
\affiliation{Pitaevskii BEC Center, CNR-INO and Dipartimento di Fisica, 
Università di Trento, I-38123 Trento, Italy }

\author{Lipeng Chen\,\orcidlink{0009-0002-1541-8912}}
\email{chenlp@zhejianglab.com}
\affiliation{Zhejiang Laboratory, Hangzhou 311100, China}

%%%%%%%%%%%%
\begin{abstract}
Describing the dynamics of strong-laser driven open quantum systems is a very challenging task that requires the solution of highly involved equations of motion. While machine learning techniques are being applied with some success to simulate the time evolution of individual quantum states, their use to approximate time-dependent operators (that can evolve various states) remains largely unexplored. In this work, we develop driven \textit{neural quantum propagators} (NQP), a universal neural network framework that solves driven-dissipative quantum dynamics by approximating propagators rather than wavefunctions or density matrices. NQP can handle arbitrary initial quantum states, adapt to various external fields, and simulate long-time dynamics, even when trained on far shorter time windows. Furthermore, by appropriately configuring the external fields, our trained NQP can be transferred to systems governed by different Hamiltonians. We demonstrate the effectiveness of our approach by studying the spin-boson and the three-state transition Gamma models.
\end{abstract}

\maketitle

Thanks to rapid advances in laser technology, resear\-chers can now manipulate quantum pathways by fine-tun\-ing the properties of strong laser fields, which allows for the amplification of specific signals that would otherwise be too weak to detect using conventional methods \cite{Jonas2003, consani2014jpca, keefer2021prl}. These developments in strong field spectroscopy provide valuable insights into complex molecular systems, significantly improving the signal-to-noise ratio and leading to more precise observations \cite{gelin2011jpcl, woerner2021jcp, egorova2008jcp, mukamel1999}. Numerical methods for solving the corresponding equations of motion (EOM) generally fall into two categories: iterative-based approaches, such as Runge-Kutta, and methods based on the time-dependent variational principle,
which minimizes the residual of the EOM solution
\cite{hoff2012pccp, norambuena2024prl, PhysRevA.109.L050201,D4DD00153B, PhysRevLett.127.230501, kramer1981geometry, PhysRevLett.122.250503}. However, despite varying levels of sophistication, these algorithms are often hindered by the high computational cost of iteratively propagating the dynamics.

The rise of machine learning has opened new avenues for theoretical simulations of chemical systems \cite{keith2021chemrev, ceriotti2021jcp, Lange_2024, Jumper2021}. Among these advancements, deep neural networks are now routinely used to approximate force fields in molecular dynamics, neural functionals in density functional theory, or potential energy surfaces in \textit{ab initio} calculations, offering significantly reduced computational costs while maintaining good accuracy \cite{noe2020, du2024jcp, unke2021chemrev, Martinetto_2024, Gedeon_2022,Nagai2020,science2021,PhysRevLett.133.098201,Gong2024}. 
Recently, there has been a growing interest in using operator learning, where neural networks act as surrogate maps for the solution operators of partial differential equations, accommodating a wide range of initial and boundary conditions \cite{Vinuesa2022,Brunton2024}. 
A notable example is the Fourier Neural Operator (FNO), originally developed to solve the Navier-Stokes equations {\cite{kovachki2021fno, li2021fno}}. Although FNO has been successfully applied to various classical continuous systems, its potential to directly solve quantum dynamics equations remains largely unexplored, and to the best of our knowledge, it has only been applied to scattering problems \cite{PhysRevD.108.L101701,PhysRevD.110.045020}.

In this Letter, we tackle the challenge of constructing neural operators for quantum dynamics and develop a \textit{Neural Quan\-tum Propagator} (NQP) model for driven-dissipative quantum systems. 
Originally designed as a solver for Mar\-kovian quantum master equations (QME), the NQP has already been applied to model population dynamics and compute various response functions for ti\-me-independent (i.e., without external driving forces) Hamiltonians \cite{zhang2024jpcl}. 
Here, we aim to overcome this limitation by designing a new architecture for NQP that incorporates external driving fields as additional input. By using the QME as the EOM for the system, our NQP acts as a universal surrogate solver applicable to arbitrary initial states and a wide range of external fields. 
Unlike conventional FNO, which assumes classical, local interactions, our concept of NQP explicitly focuses on quantum dynamics involving thus non-local quantum effects. 
Notably, due to its rigorous adherence to the composition property of quantum propagators (an aspect rarely explored in the literature of operator learning \cite{McGreivy2024,alkin2024upt,JMLR:v24:21-1524}), NQP can predict long-time dynamics, far beyond the training time window. Moreover, we demonstrate that
our model can be easily transferred to various systems governed by different Hamiltonians.

The remainder of the paper is organized as follows. First, we present the QME for driven-dissipative quantum systems along with the key technical details of our NQP architecture. We then discuss the training scheme and test the NQP numerical performance by using the spin-boson and the three-state Gamma models. Finally, we offer some conclusions.

\textit{Driven-dissipative quantum dynamics.---} In this work, we are interested in open quantum systems driven by external fields. The Hamiltonian is written as
\begin{equation}
\hat{H}(t) = \hat{H}_{0} + \sum_{k=1}^K f_{k}(t) \, \hat{F}_{k}\,,
\label{eq.H_total}
\end{equation}
where the first term is the system Hamiltonian:
\begin{equation}
\hat{H}_{0} = \sum_{j} \varepsilon_{j} |j\bigr\rangle \bigl\langle j| + 
\sum_{j \ne j^{\prime}} \Delta_{j j^{\prime}} |j\bigr\rangle \bigl\langle j^{\prime}|,
\end{equation}
with $\varepsilon_{j}$ being the state energies and $\Delta_{j j^{\prime}}$ the interstate couplings. In the second term, $f_{k}(t)$ is a scalar function and $\hat{F}_{k}$, the operator part of the $k$-th external field. The overall system is further interacted with a set of Markovian heat baths and the EOM is described by the QME {\cite{weiss2012, breuer2007, PRXQuantum.5.020202}}:
\begin{equation}
\partial_t \hat{\rho}(t) = - \frac{i}{\hbar} \big[ \hat{H}(t), \, \hat{\rho}(t) \big]  
- \sum_{j} \gamma_{j} \mathcal{D}_{j} \left[\hat{\rho}(t) \right] ,
\label{eq.qme}
\end{equation}
where $\hat{\rho}(t)$ is the density operator of the system, and 
\begin{equation}
\mathcal{D}_{j} \left[\hat{\rho}(t) \right] = \left( \hat{V}_{j}^{\dagger} \hat{V}_{j} \hat{\rho}(t)  
+ \hat{\rho}(t) \hat{V}_{j}^{\dagger} \hat{V}_{j} - 
2 \, \hat{V}_{j} \hat{\rho}(t) \hat{V}_{j}^{\dagger} \right)
\end{equation}
with $\gamma_{j}$ and $\hat{V}_{j}$ being the coupling strength and the operator function of the $j$-th heat bath, respectively. In the presence of an external field, the time-dependent propagator can be defined through the equation:
\begin{equation}
\hat{\rho}(t) = \mathcal{T}_{+} \exp \left[ \int_{t_0}^{t} {\rm{d}} s \, \mathcal{L}(s) \right] \hat{\rho}(t_0),
\label{eq.prop}
\end{equation}
where $\mathcal{T}_{+}$ denotes the time-ordered operator and the su\-per-operator $\mathcal{L}(s)$ corresponds to the right-hand side of Eq.~{\eqref{eq.qme}}. 

We now define a discrete time grid with a time step $\delta_t$ and $t_n = n \delta_t$.
The $k$-th external field at $t_n$ is denoted by $f_{n}^{k} = f_k(t_n)$.
All $K$ external fields from $t_m$ to $t_{n}$ can thus be represented as a matrix, 
\begin{equation}
{\vec{f}}_{n, m} = 
\begin{bmatrix}
& f_{m}^{1} & f_{m}^{2} & \cdots & f_{m}^{K} \\
& f_{m+1}^{1} & f_{m+1}^{2} & \cdots & f_{m+1}^{K}\\
& \vdots & \vdots & \ddots & \vdots \\
& f_{n}^{1} & f_{n}^{2} & \cdots & f_{n}^{K}
\end{bmatrix} \,,
\end{equation}
which is further aligned as a vector and used as the input of the neural network.

Next, we introduce the collective index $x = (j, j^{\prime})$ and align the density operator as a vector, $\hat{\rho}(t_n) \sim \vec{\rho}_n = \{ \rho(x_0, t_n), \rho(x_1, t_n), \cdots \}$, with $\rho(x, t) = \langle j | \hat{\rho}(t)| j^{\prime} \rangle$.
Formally this vectorization is identical to using the twin-space or Liouville space representation \cite{10.1063/1.5115323}. The propagator defined in Eq.~\eqref{eq.prop} can be recast into the matrix-vector form as
\begin{equation}
\vec{\rho}_{n} = {\bm{G}}\left[{\vec{f}}_{n, m}; \vec{\rho}_m \right].
\label{eq.prop_mv}
\end{equation}
Here the propagator ${\bm{G}}$ can be regarded as a functional that projects the given initial state $\vec{\rho}_m$ and the external field $\vec{f}_{n, m}$ to their corresponding final state $\vec{\rho}_n$ satisfying Eq.~\eqref{eq.qme}. As a consequence of the linearity of quantum mechanics, the composition property holds and reads
\begin{equation}
\vec{\rho}_{n_2} = {\bm{G}}\left[{\vec{f}}_{n_2, n_1}; 
{\bm{G}}\left[{\vec{f}}_{n_1, n_0}; \vec{\rho}_{n_0} \right] \right].
\label{eq.prop_comp}
\end{equation}
Although our discussion is limited to the Markovian assumption, the same protocol can be easily extended to non-Markovian cases by replacing QME with other numerical exact EOM such as the hierarchical equations of motion \cite{zhang2024es}.

\begin{figure}[t!]
\centering
\includegraphics[width=0.5\textwidth]{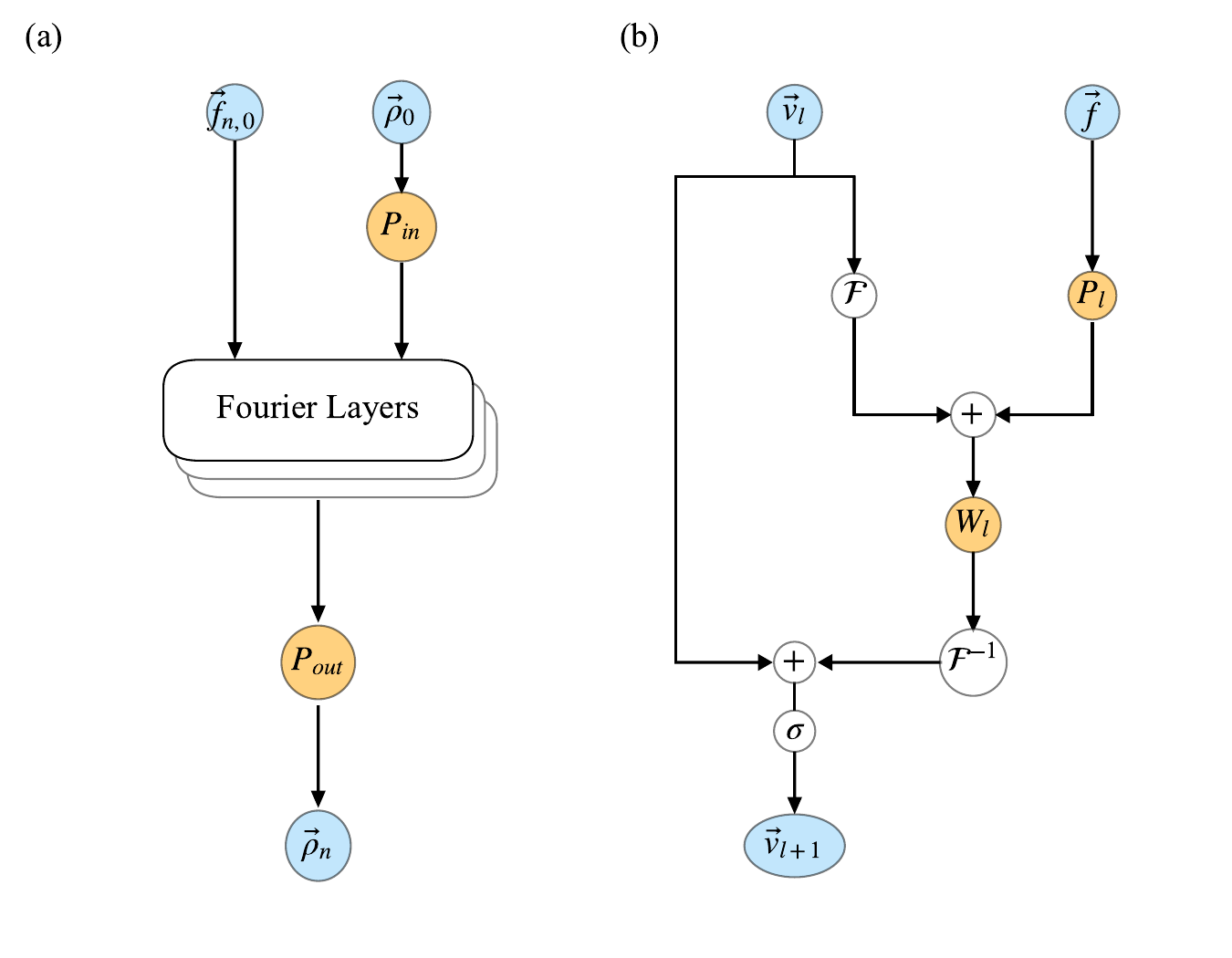}
\caption{The architecture of (a) the driven NQP model, and (b) the $l$-th Fourier layer, respectively. $\mathcal{F}$ and $\mathcal{F}^{-1}$ denote the Fourier transform and its inverse. $+$ and $\sigma$ are the element-wise addition and the GeLU activation function.}
\label{fig.model}
\end{figure} 

%%%%%%%%%%
\textit{Neural Quantum Propagators with driven fields.---} To construct the NQP for driven dynamics, we follow the adaptive FNO transformer {\cite{guibas2021afno}} and stochastic Fourier Differential Equation {\cite{salvi2021spde}} architectures. Our model thus takes an arbitrary initial state $\vec{\rho}_{0}$ and external field $\vec{f}_{n, 0}$ as the input, and outputs the target $\vec{\rho}_{n}$, following Eq.~\eqref{eq.prop_mv}. The upper time limit of the model is chosen as $ n \le N_t$ $(N_t = t_{max}/\delta_t)$. For ease of numerical construction, we limit $n$ to be an integer but higher resolution can be obtained by either reducing the time step or adopting the super-resolution algorithm. Crucially, because of the composition property of quantum propagators, long-time dynamics outside this time interval can be obtained by recursively applying the propagator. 

Fig.~\ref{fig.model} sketches the architecture of our driven NQP model. In panel (a), $P_{in}$ and $P_{out}$ are linear feedforward neural networks that serve as the projection between the physical space and the latent Fourier space. As shown in panel (b), the $l$-th Fourier layer performs the following ope\-ration:
\begin{equation}
\vec{v}_{l+1} = \sigma \left( \vec{v}_{l} + \mathcal{F}^{-1} \left[ {\bm{W}}_{l} 
\left(\mathcal{F}[\vec{v}_{l}] + {\bm{P}}_{l} {\vec{f}}_{n,0}\right)\right]\right).
\end{equation}
The model first performs the Fourier transform on $v_{l}$, which is the output of the previous layer or $P_{in}$.
It then passes the external field vector $\vec{f}_{n, 0}$ to another linear feedforward neural network $P_{l}$.
The result of these two routes is then added together and passed to the point-wise convolution $W_{l}$, which serves as learnable parameters on the Fourier space.
It is then followed by the inverse Fourier transform and nonlinear Gaussian Error Linear Unit (GeLU) activation function.
The left-most route serves as the residual connection, which is introduced to improve numerical stability.
The learnable parameters are all those contained in $P_{in}$, $P_{out}$, $W_{l}$, and $P_{l}$, respectively.

The key novelty of our model can be further appreciated by the following fact: removing the right-most route associated with $P_{l}$ results in the (original) non-driven NQP \cite{zhang2024jpcl}, as sketched in Fig.~{\ref{fig.model}}(b). The inclusion of $\vec{f}_{n,0}$ introduces the correction to $\vec{v}_{l}$ in Fourier space that comes from the effects of the external field. For simplicity, we use a linear layer for $P_{l}$ and merge with $\mathcal{F}[\vec{v}_{l}]$ through the element-wise addition. Replacing them with more specific structures, such as convolution, may improve the performance, but will not be discussed in this work.

%%%%%%%%%%
\textit{Physics-informed training algorithm.---} To train the NQP model, we adopt the physics-informed training algorithm and define the loss function
\begin{equation}
L = \alpha L_{data} + (1-\alpha)L_{phys}\,,
\label{eq.loss}
\end{equation}
where $L_{data}$ and $L_{phys}$ are the data and physics-informed loss functions, and $\alpha$ is a hyper-parameter that will be dynamically adjusted during the training process.

The first term $L_{data}$ is evaluated from a prepared training dataset, which contains in total $N_{data}$ samples. The $p$-th sample of the dataset is composed of $\vec{f}^{(p)}_{n,0}$, an external field vector, and $\vec{\rho}_{n}^{(p)}$, the corresponding dynamical states. $L_{data}$ is thus defined as
\begin{equation}
L_{data} = \frac{1}{N_{data} N_t} \sum_{p=1}^{N_{data}} \sum_{n=0}^{N_t} 
{\left| \left|  \vec{\mu}_{n}^{(p)} - \vec{\rho}_{n}^{(p)} \right|\right|}_{F},
\end{equation}
where $|| \cdot ||_{F}$ denotes the Frobenius norm, $\vec{\mu}_{n}^{(p)}$ is the dynamical state for the $p$-th sample predicted by the model via the equation
\begin{equation}
\vec{\mu}_{n}^{(p)}  = {\bm{G}}_{\theta}\left[ \vec{f}_{n,0}^{(p)}; \vec{\rho}_0^{(p)}\right],
\label{eq12}
\end{equation}
where ${\bm{G}}_{\theta}$ is the aforementioned NQP model (with $\theta$ being the entire set of learnable parameters) and $\vec{\rho}_0^{(p)}$ is the initial state for the $p$-th sample.
This training dataset is pre\-pa\-red by solving the time evolution using some conventional iterative method.

The second term $L_{phys}$ is defined as the residual of Eq.~{\eqref{eq.qme}} \cite{Karniadakis2021}:
\begin{equation}
\label{physin}
L_{phys} = \frac{1}{N_{phys} N_t}  \sum_{p=1}^{N_{phys}} \sum_{n=0}^{N_t} 
{\left| \left|  \partial_t\vec{\mu}_{n}^{(p)} - {\bm{\mathcal L}}_{n}\vec{\mu}_{n}^{(p)} \right|\right|}_{F},
\end{equation}
where ${\bm{\mathcal L}}_{n}$ represents the right-hand side (time-de\-pen\-dent) operator of Eq.~\eqref{eq.qme}. To evaluate such a cost function, we have introduced a physics dataset with in total $N_{phys}$ samples. Interestingly, compared to $L_{data}$, the evaluation of $L_{phys}$ requires only the initial states. In fact, the explicit expression of each dynamical state $\vec{\mu}_{n}^{(p)}$ in Eq.~\eqref{physin} can be avoided by using the predicted propagator ${\bm{G}}_{\theta}$ that relates this state with the initial one $\vec{\rho}_{0} \sim \hat{\rho}(0)$, as indicated in Eq.~\eqref{eq12}. %As a result, rather large $N_{phys}$ can be adopted in the training stage. 

We now present numerical results for the spin-boson system and three-state transition Gamma system. The accuracy of the model is demonstrated by comparing the time evolution of the density operator with the result from the fourth-order Runge-Kutta method (RK4).

%%%%%%%%%%%%
\textit{Numerical experiments.---} The spin-boson system, relevant for quantum control of pulse reverse engineering and controllable dissipative dynamics {\cite{norambuena2024prl, ran2020pra, wu2021scirep}}, is governed by the Hamiltonian:
\begin{equation}
\hat{H}_{0} = \frac{\omega_{z}}{2} \left( |e\bigr\rangle \bigl\langle e| - |g\bigr\rangle \bigl\langle g| \right)
+ \omega_{x} \left( |e\bigr\rangle \bigl\langle g| + |g\bigr\rangle \bigl\langle e| \right).
\end{equation}
We assume two heat baths: $\hat{V}_{1} =  |e\bigr\rangle \bigl\langle g|$ and $\hat{V}_{2} = |g\bigr\rangle \bigl\langle e|$, describing the absorption and emission processes. For the external field, we assume only one field, with $\hat{F} = |e\bigr\rangle \bigl\langle e|$, and $f(t)  = e^{i \omega_{f} t}$. In what follows, we choose $\omega_{z} = \omega_{0}$ and use it as the unit for all the other parameters. The other parameters are chosen as $\omega_{x} = 0.5$, $\gamma_{1} = 0.1$ and $\gamma_{2} = 0.2$, respectively. 
The range of $\omega_{f}$ is chosen as $\omega_{f} \in (0.2, 1.0)$. For the time parameters, we let $\delta_t = 0.05$. The time limit $t_{max}$, associated with time point $N_{t}$, plays an important role in the performance of the model and is discussed in the Appendix. We sample the initial states to evaluate $L_{phys}$ \eqref{physin} from the Gaussian Unitary Ensemble as explained in more detail also in the Appendix.

%%%%%%%%%%
\begin{figure}
\centering
\includegraphics[width=0.5\textwidth]{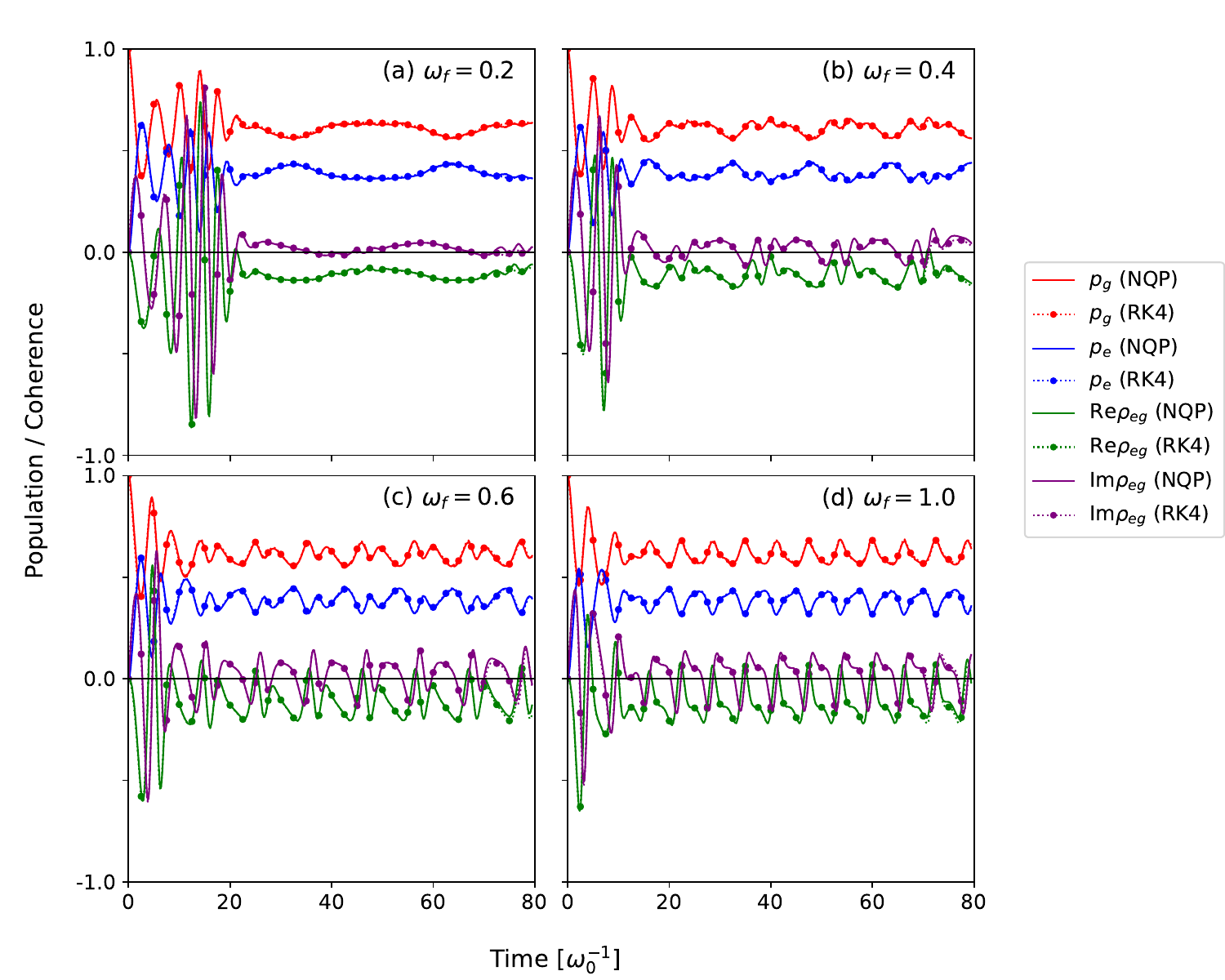}
\caption{The population and coherence dynamics of the spin-boson model at $\omega_{f} = $ (a) $0.2$, (b) $0.4$, (c) $0.6$, and (d) $1.0$ as predicted by the NQP model. The red, blue, green, and purple curves represent the populations, $p_{g}$, $p_{e}$, and the real and imaginary parts of $\rho_{eg}$, respectively. The solid and dashed lines represent the results from the NQP model and the reference RK4 (shown with a marker for better illustration), respectively.}
\label{fig.spin_20_diag}
\end{figure} 
%%%%%%%%%%

We choose $t_{max} = 20$ with $N_t = 400$ and the initial state $\hat{\rho}(0) = | g \rangle \langle g |$.
We focus on the time evolution of the density operator up to $t=80$ by comparing it with the reference result obtained from the RK4 using a time step of  $0.05$.
In Fig.~\ref{fig.spin_20_diag}, we present the dynamics of populations $p_{j}(t) = \langle j | \hat{\rho}(t)| j\rangle$ $(j = g, e)$ and coherence $\rho_{eg}(t) = \langle e| \hat{\rho}(t) | g\rangle$ at $\omega_{f} = $ (a) $0.2$, (b) $0.4$, (c) $0.6$, and (d) $1.0$, respectively, representing the typical slow, intermediate, and fast modulation forces.
As shown in Fig.~{\ref{fig.spin_20_diag}}, the system exhibits different dynamics within $t \le 20$ at these four typical $\omega_{f}$ cases. Notice that our model can not only predict the time evolution up to $t_{max}$ with high accuracy but also produce the correct long-time coherent dynamics for time far beyond $t_{max}$.

\textit{Universality.---} One of the main features of the NQP introduced in this Letter is its universality, which renders it applicable to different Hamiltonians. This is established by assuming the constant force, $f_{k}(t) = c_{k}$, with ${c}_{k} \in (c_{k, \rm{min}}, \, c_{k, \rm{max}})$. 
The effective system Hamiltonian now becomes $\hat{H}_{c} = \hat{H}_{0} + \sum_{k}^{K} {c}_{k} \hat{F}_{k}$. Our trained model can be easily employed to predict dynamics for all $\hat{H}_{c}$. This significantly improves the transferability over previously developed models, which, being limited to a specific system, have to be re-trained when applied to a different one \cite{zhang2024jpcl}. To test the performance of our NQP model when applied to different Hamiltonians, we adopt, in the following, the three-state transition Gamma model.

\textit{Three-state Gamma system.---} This model plays an important role in stimulated Raman adiabatic population transfer \cite{kuklinski1989pra, bergmann1998rmp}.
The system Hamiltonian is defined as
\begin{equation}
\begin{aligned}
\hat{H}_{c_1, c_3} &= \sum_{j=1}^{3} \omega_{j} \bigl| j \bigr\rangle \bigl \langle j \bigr| + 
c_{1} \left( \bigl| 1 \bigr\rangle \bigl \langle 2 \bigr| + \bigl| 2 \bigr\rangle \bigl \langle 1 \bigr| \right) \\
&+ c_{3} \left( \bigl| 2 \bigr\rangle \bigl \langle 3 \bigr|  +  \bigl| 2 \bigr\rangle \bigl \langle 3 \bigr| \right).
\end{aligned}
\label{eq.three_state}
\end{equation}
We treat the first term as $\hat{H}_0$ and the rest as constant external fields. The state $|2\rangle$ is the transition state between $|1\rangle$ and $|3\rangle$. The transition between different states can be tuned by varying the interstate couplings $c_{1}$ and $c_{3}$. 
The system-bath coupling operator is chosen as $\hat{V}_{j} = |j\rangle\langle j|$ for $j = 1 \sim 3$, respectively.
We set the energies of three states as $\omega_{1} = 0.0$, $\omega_{2} = 0.1$, and $\omega_{3} = 1.0$. For the heat baths, we set $\gamma_{1} = \gamma_{3} = 0.1$, $\gamma_{2} = 0.2$, assuming a stronger dissipation for the transition state. The range of $c_{1}, c_{3}$ is chosen as $c_{1}, c_{3} \in (0.2, ~ 0.8)$.

%%%%%%%%%%
\begin{figure}[t]
\centering
\includegraphics[width=0.5\textwidth]{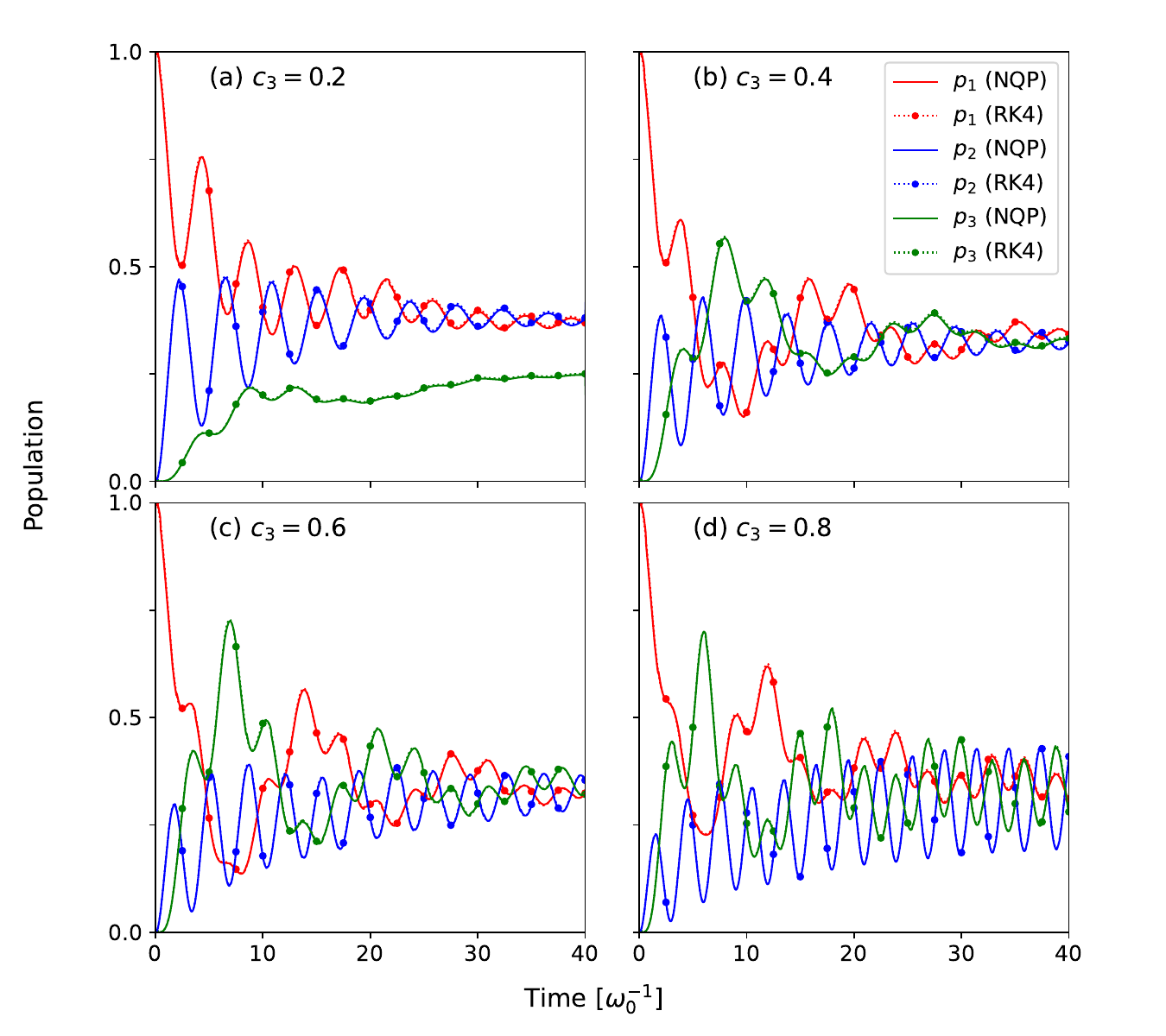}
\caption{The population dynamics of the three-state system at $c_1 = 0.3$ and different $c_3$ cases.
The solid line represents the results evaluated from our NQP model, which was shared among all the cases. The dotted line represents the reference RK4 result.}
\label{fig.three_case_1}
\end{figure} 
%%%%%%%%%%

We choose $t_{max} = 2$, $\delta_t = 0.05$, and the initial state  $\hat{\rho}(0) = |1\rangle\langle 1| $. In Fig.~{\ref{fig.three_case_1}}, we show population dynamics up to $t = 40$ for $c_{1} = 0.3$ and $c_{3} = $ (a) 0.2, (b) 0.4, (c) 0.6, and (d) 0.8, respectively. For all cases, our NQP model predicts accurate dynamics at times far beyond $t_{max}$, demonstrating the adaptability of the model on different Hamiltonians. In addition, we found that changing $t_{max}$ in the range of $0.5 \sim 10$ does not show any apparent differences in the performance of the model. This suggests that different Hamiltonians can be easily handled by our NQP model, even with small training time windows. 

%%%

%%%%%%%%%%%%
\textit{Conclusion.---} In summary, we developed an NQP mo\-del ---a universal neural network architecture that treats external fields as additional inputs and can handle arbitrary initial quantum states--- for simulating driven-dissi\-pa\-tive quantum dynamics. It is striking that the trained NQP can predict quite a long-time dynamics for different external fields far beyond the training time window. We also showed that, by appropriately configuring the external fields, the model can be transferred to systems described by diffe\-rent Hamiltonians. These results highlight the flexibility and expressibility of our NQP model, rendering it a power\-ful tool for studying quantum dynamical problems.

One of the main issues of the NQP model is that the number of learnable parameters scales exponentially with the system size \cite{Brunton2024}. Similar to other operator-learning frameworks, this scaling problem arises from the fact that the model approximates a functional of inputs with high dimensionality {\cite{kovachki2021fno}}. To reduce computational costs in future applications, one potential solution is to represent the density matrix using tensor network structures, which are known for their polynomial scaling \cite{cirac2021rmp, ren2022wcms, li2020jcp, borrelli2021wcms, novikov2015tnn}. Alternatively, insights can be drawn from exact expressions for steady states \cite{PhysRevX.10.021022}. Finally, another promising avenue of research is to reverse the learning protocol, using the learned propagators for quantum control problems, crucial for the development of quantum technologies \cite{Brif_2010, PhysRevApplied.17.064028}. 

%%%%%%%%%%%%%
\section*{Acknowledgments}
We thank Maxim Gelin for enlightening discussions during the preparation of the manuscript. J.Z. and L.P.C. acknowledge support via a starting grant of the research center of new materials computing of Zhejiang Lab (3700-32601). C.L.B.-R.~gratefully acknowledges financial support from the Royal Society of Chemistry and the European Union’s Horizon Europe Re\-search and Innovation program  un\-der the Marie Skło\-dowska-Curie Grant Agreement n°101065295--RDMFTforbosons. Views and opinions expressed are however those of the author only and do not necessarily reflect those of the European Union or the European Research Executive Agency.

%\printbibliography

\section{Appendix}

\textit{Preparation of the training set.---}
We briefly introduce the preparation of data and physics set used in our experiments. The initial state $\vec{\rho}_{0} \sim \hat{\rho}(0)$ is randomly sampled from the Gaussian Unitary Ensemble as Hermitian matrix.
A sigmoid scaling is conducted for the diagonal entries to have trace-norm,
\begin{equation}
\bigl\langle j \bigr| \hat{\rho} \bigl| j \bigr\rangle \to 
\frac{\bigl\langle j \bigr| \hat{\rho} \bigl| j \bigr\rangle }
{\left( \sum_{j} \bigl\langle j \bigr| \hat{\rho} \bigl| j \bigr\rangle \right)}.
\end{equation}
The external field vector $\vec{f}_{n,0}$ is prepared by first selecting a specific function form for $f(t)$, and then randomly choosing the related parameters within chosen intervals.
For better understanding, we use periodic forces as  $f(t) = \exp(i \omega_{f} t)$.
The random sampling is conducted by first choosing $\omega_{f}$ as a uniform random number within the range $\omega_{f} \in (\omega_{\rm{min}}, ~ \omega_{\rm{max}})$. Finally, the vector $\vec{f}_{n,0}$ is obtained by embedding $f(t)$ to the time grid $t_{n}$.

\textit{Computational Details.---}
The hyper-parameters and training setup of the NQP model are chosen as follows. $P_{in}$, $P_{out}$ and $P_{l}$ are parameterized as 2-layer neural networks with a hidden channel of size 256. For $W_{l}$, we use 4 Fourier layers, each of which has a hidden channel of size 128. All the Fourier modes are explicitly included in the model. The total number of trainable parameters is around 2 millions. We prepare the training dataset by randomly sampling $N_{data} = 2000$ initial states and external field parameters, and integrating the EOM with the RK4. The physics dataset has a size of $N_{phys} = 200$, and is regenerated at each epoch. The model is trained by optimizing Eq.~10 (see main text) with $\alpha = 0.5$ using the Adam optimizer and a learning rate of $10^{-4}$. The training is conducted for $10^{4}$ epochs on a single Nvidia 4090 GPU card within $5$ hour until the loss function reaches $\sim 10^{-4}$. The trained model can then be applied to any initial states and any $\omega_{f} \in (0.2, 1.0)$.

%%%%%%%%%%
\begin{figure}
\centering
\includegraphics[width=0.5\textwidth]{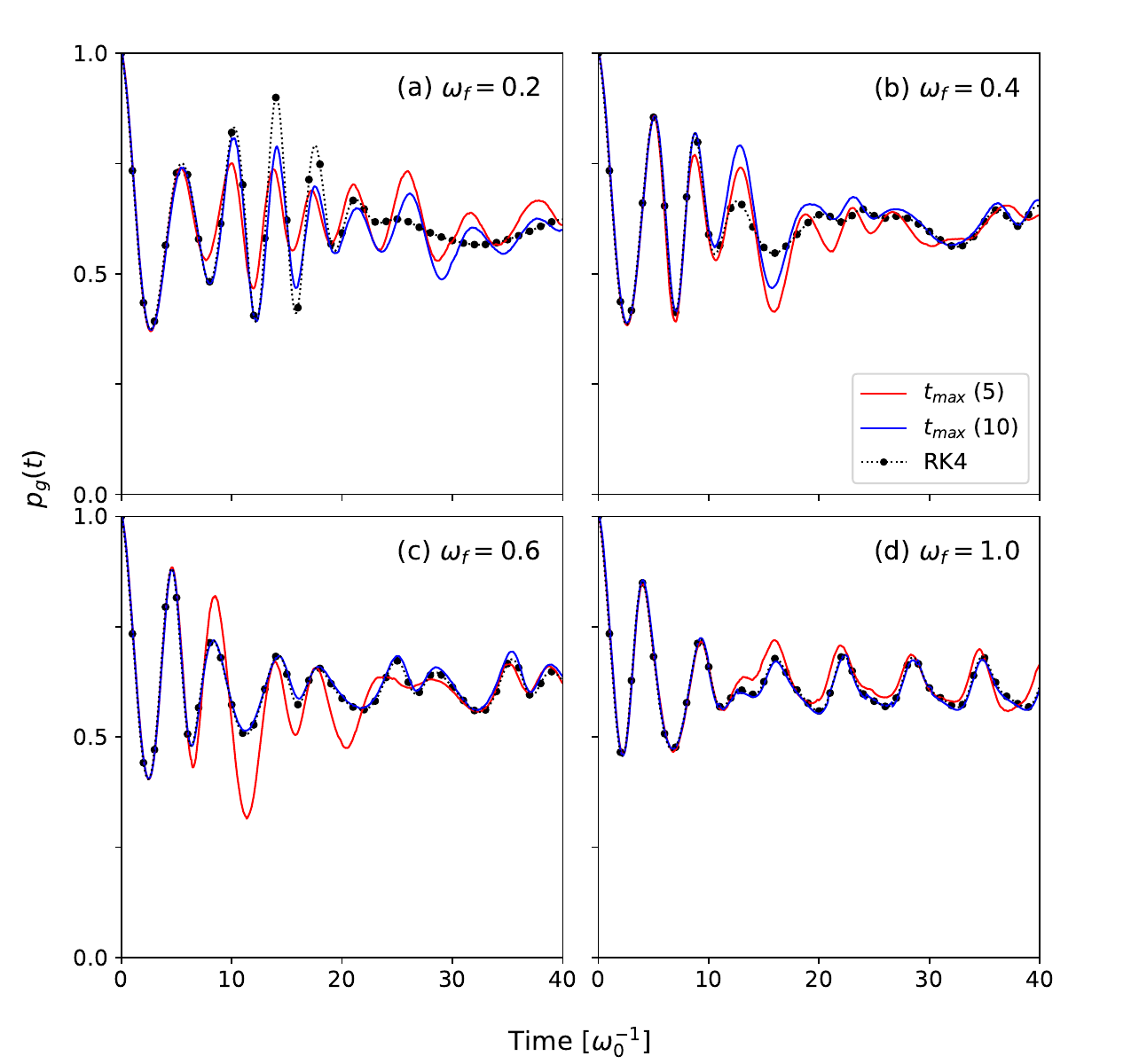}
\caption{The population dynamics $p_{g}(t)$ for different $\omega_{f}$ evaluated from the NQP model with $t_{max} = 5$ (red) and $10$ (blue). The reference RK4 results are shown in black.}
\label{fig.spin_diff}
\end{figure} 
%%%%%%%%%%
\textit{Performance outside the training window.---}
We note that the performance of the model strongly depends on the choice of the maximum time $t_{max}$, which should be, in principle, large enough. To test the effect of $t_{max}$ on the model's performance, we retrained the model using smaller $t_{max}$. 
We show the population dynamics $p_{g}(t)$ up to $t=40$ evaluated from the NQP model with $t_{max} = 5$ and $10$ in Fig.~{\ref{fig.spin_diff}}. The other settings are the same as the case of $t_{max} = 20$ in Fig.~2 (see main text). As shown in Fig.~{\ref{fig.spin_diff}}, the trained models can always predict the correct dynamics within the training time window $t_{max}$. For $t_{max} = 5$, the model incorrectly predict  dynamics at $t > t_{max}$ for all $\omega_{f}$. We found that employing a larger model by increasing the number of layers to 15 and channels to 512 merely leads to longer training time, but cannot improve long-time dynamics outside the training window. Increasing $t_{max}$ to 10, the performance of the model improves with the increase of $\omega_f$, yielding almost correct dynamics for the cases of $\omega_f=0.6$ and 1.0. It is thus expected that the optimal choice of $t_{max}$ is determined by the range of $\omega_f$, which corresponds to the modulation of the external fields. This hyper-parameter has to be carefully chosen in order to get better performance.

\bibliography{ref_list}

\end{document}